\documentclass[10pt, conference, compsocconf]{IEEEtran}
\ifCLASSINFOpdf
\else
\fi
\usepackage{caption}
\usepackage{subfigure}
\usepackage{amsthm}

\usepackage{amsmath}
\usepackage{mathtools}
\usepackage{url}
\usepackage{algorithm}
\usepackage{algorithmic}
\usepackage{dsfont}
\usepackage{bm}

\usepackage{cite}
\usepackage{amsmath,amssymb,amsfonts}
\usepackage{algorithmic}
\usepackage{graphicx}
\usepackage{textcomp}
\usepackage{xcolor}

\usepackage{booktabs}
\usepackage{multirow}
\usepackage{makecell}
\usepackage{adjustbox}

\begin{document}
%



%

\title{ Tiansuan Constellation: An Open Research Platform}

\author{\IEEEauthorblockN{ Shangguang Wang, Qing Li,  Mengwei Xu,  Xiao Ma, Ao Zhou, Qibo Sun}
\IEEEauthorblockA{\textit{State Key Laboratory of Networking and Switching Technology} \\
\textit{Beijing University of Posts and Telecommunications}\\
Beijing, China \\
\{sgwang;q\_li;mwx;maxiao18;aozhou;qbsun\}@bupt.edu.cn}
}
\maketitle

\begin{abstract}

Satellite network is the first step of  interstellar voyages. It can provide global Internet connectivity  everywhere on  earth, where most areas  cannot access the Internet  by the terrestrial infrastructure  due to the geographic accessibility and high cost. 
The space industry experiences  a  rise in large low-earth-orbit satellite constellations to achieve universal connectivity.  The research community is also urgent to do some leading research to bridge the connectivity divide. Researchers now conduct their work by simulation, which is far from enough.  However, experiments on real satellites are blocked  by the high threshold of space technology, such as deployment cost and  unknown risks.
To solve the above dilemma, we are eager to  contribute to the  universal connectivity and  build an open research platform, Tiansuan constellation to support experiments on  real satellite networks. We discuss the potential research topics that would benefit from Tiansuan constellation. We provide
two case studies that have already deployed in  two  experimental satellites of Tiansuan constellation.

\end{abstract}
\IEEEpeerreviewmaketitle
\section{Introduction}



Human society is facing development bottlenecks such as  global warming, prominent social contradictions, the coexistence of industrial surplus, and resource depletion. Would human need interstellar voyages and expand the boundaries of human existence like the age of discovery? If yes, the satellite network will play a major role at the beginning of the era of interstellar voyages.
Moreover, there  still be more than 80\% of the land area and more than 95\% of the ocean area without network access $\footnote{https://www.ccidgroup.com/info/1096/21569.htm}$.
It is impossible  for the terrestrial network to provide ubiquitous broadband internet access because of geographic accessibility and  the high cost of  building infrastructure.

 Low Earth Orbit (LEO) satellites are promising to provide global internet and service due to the reduced cost of satellite component manufacturing and  ride-sharing launch. Multiple space companies are gearing up to deploy  LEO constellations  to provide global  low-latency high-bandwidth  Internet.  Dated to November 2021, SpaceX has now launched 1,844 Starlinks and London-based OneWeb has launched 358 internet satellites.   Besides, Microsoft and Amazon offer  “Ground station as a service” for space customers $\footnote{https://aws.amazon.com/cn/ground-station/}$.

 This exciting development is taking shape rapidly in the industry, while the  research community is urgent to do some leading research to bridge the connectivity divide.  Large-scale constellations   feature hundreds or  thousands of low-volume satellites, each with an orbital period of ~100 minutes. This new type of  infrastructure brings  inherent challenges. Recent work has highlighted the challenges the  dynamic connectivity could create at various fields such as network (e.g., topology \cite{bhattacherjee2019network}, routing\cite{giuliari2020internet}, congestion control \cite{kassing2020exploring}),  earth observation \cite{vasisht2021l2d2},  in-orbit edge computing \cite{10.1145/3422604.3425937}. While it is useful to explore these problems using simulation tools, ultimately,   we would like to conduct experiment evaluations on the real satellite constellation.  However,  due to the high technical threshold of space technology, such as deployment cost,  unknown risks, most scientific research researches are blocked.  SatNetLab calls researchers to  conduct some network experiments on commercial satellite Internet as users \cite{10.1145/3464994.3465000}, while experiments changing the network settings or to be deployed on satellites may be  not allowed. 
  \begin{figure}[t]
\begin {center}
\centerline{\includegraphics[width=0.4\textwidth]{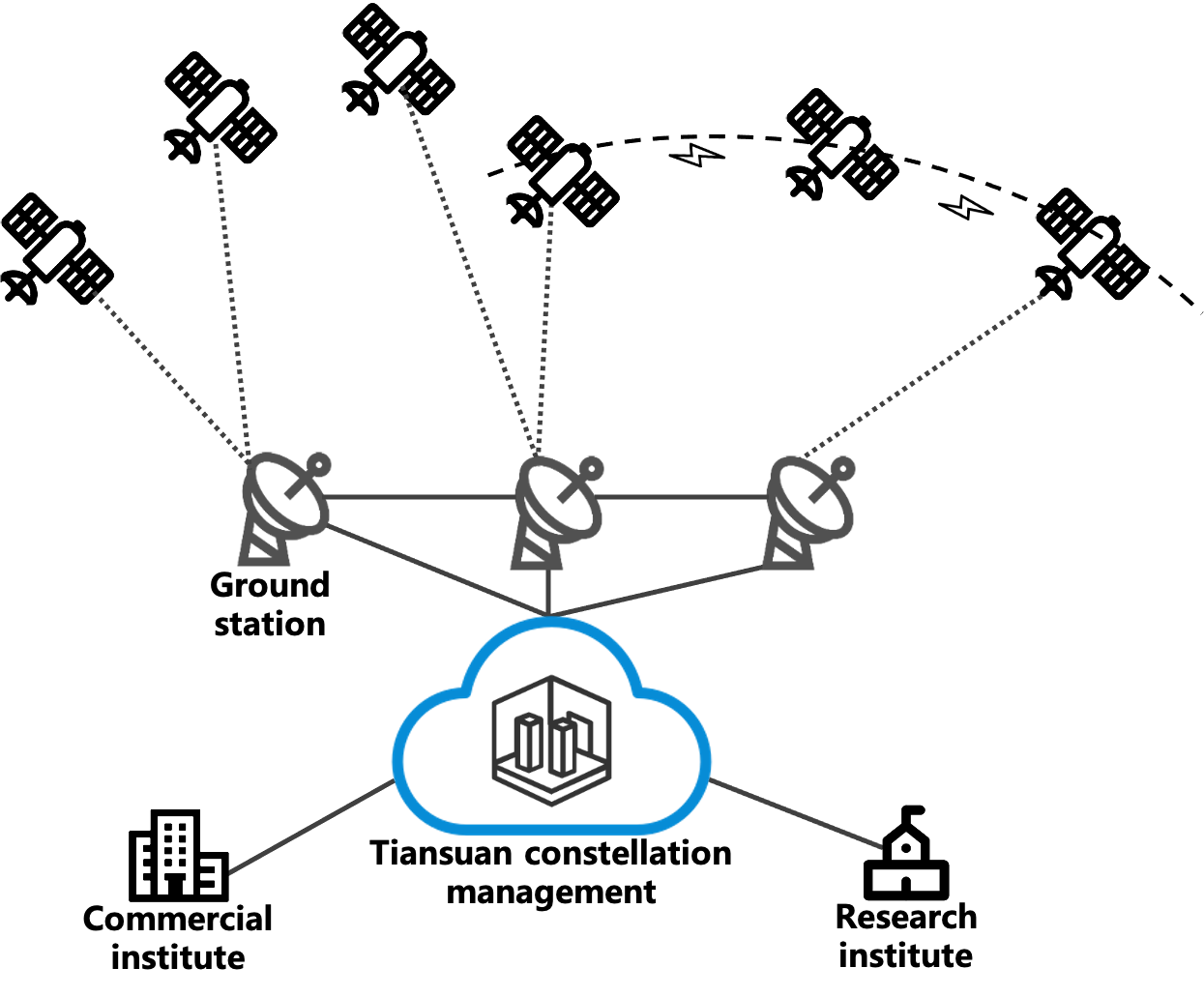}}
\caption{ Tiansuan Constellation Framework.}
\label{Fig:system}
\end {center}
\vspace{-6ex}
\end{figure}

To address these limitations, we are eager to contribute to the  universal connectivity and  propose Tiansuan constellation, an open research platform  illustrated  in Fig.~\ref{Fig:system}.  Tiansuan constellation plans three stages, the first stage with 6 satellites,  the second stage with 24 satellites, and  the third stage with stage 300 satellites. Except for satellites owned by our institute,  satellites carrying our payloads and satellites joining our Tiansuan plan together form the 
 Tiansuan constellation.   This global research platform would enable 1) experiments on  real satellites and 2)  testing practical solutions to improve the  performance of the constellation.  Thus, it helps studying the  impacts of the real physical environment (e.g., weather conditions), which is hard to fully capture in simulations. Moreover,  real experiments on satellite constellations can  uncover any pitfalls to the solutions for  commercial deployment potential. 

 \begin{table*}[htbp]
\renewcommand{\arraystretch}{1.5}
\caption{Main parameters of  Tiansuan  Constellation Stage 1.}
\label{sat-parameters}
\centering
\begin{tabular}{@{}c|c|c|c|c|c|c|c|c@{}}
\hline
\textbf{Number} & \textbf{Orbital Altitude} & \textbf{Mass} & \textbf{Battery Capacity} & \textbf{Spectrum} & \textbf{Uplink Rate} & \textbf{Downlink Rate} & \textbf{ISLs} & \textbf{Processors}\\ \hline 
1 & 500$\pm$50km & $\leq$ 30kg & 118Wh -- 236Wh & X-band & 0.1Mbps -- 1Mbps & 100Mbps -- 600Mbps & NO & CPU/NPU \\ \hline
2 & 500$\pm$50km & $\leq$ 30kg & 118Wh -- 236Wh & X-band & 0.1Mbps -- 1Mbps & 100Mbps -- 600Mbps & NO & CPU/NPU \\ \hline
3 & 500$\pm$50km & $\leq$ 30kg & 118Wh -- 236Wh & X-band & 0.1Mbps -- 1Mbps & 100Mbps -- 600Mbps & NO & CPU/NPU \\ \hline
4 & $>$ 500km & $>$ 50kg & $>$ 360Wh & X, Ku, Ka bands & $\ge$ 200Mbps & $\ge$ 1Gbps & YES & CPU/NPU/GPU \\ \hline
5 & $>$ 500km & $>$ 50kg & $>$ 360Wh & X, Ku, Ka bands & $\ge$ 200Mbps & $\ge$ 1Gbps & YES & CPU/NPU/GPU \\ \hline
6 & $>$ 500km & $>$ 50kg & $>$ 360Wh & X, Ku, Ka bands & $\ge$ 200Mbps & $\ge$ 1Gbps & YES & CPU/NPU/GPU \\ \hline
\end{tabular}
\end{table*}
  In the following, we introduce Tiansuan constellation design and how it could help
the  research community and LEO satellite industry in Section 2, discuss research directions Tiansuan could  support in Section 3, and provide case studies deployed on Tiansuan  constellation in Section 4. We conclude in Section 5.



\section{Tiansuan Constellation Design }

\subsection{Overview}

Tiansuan constellation  aims  to build an open platform that supports experiments including but not limited to 6G core network system, satellite Internet (Satellite Digital Networking), satellite operating system,  federated learning and AI acceleration, and on-board service capability opening. It plans three stages, the first stage with 6  satellites,  the second stage with 24 satellites, and  the third stage with 300 satellites.  In all three stages of Tiansuan constellation, there are  three satellite types.
The first type  satellites  belong to  our institute.  The  second type  of satellites  are owned by others institutes and carry our payloads which can collaborate with other payloads on the computing platform. The third type  of satellites also belong to others but  can join our open computing platform through a unified interface. 

The first satellite of Tiansuan constellation will be launched in May 2022.  Satellites in Tiansuan are manufactured  according to the standard \cite{cubeSat} and the parameters of satellites in the first stage are in Table I. Most satellites will be launched into sun-synchronous orbit. The first three satellites combine edge computing capabilities with remote sensing applications. The last three satellites are mainly to explore communication capabilities with  inter-satellite links.  Control functions are provided by the onboard computer while the majority of the computing power is provided by the payloads  as listed in Table I. 
As shown in Fig. 1, ground stations receive data from satellites and  distribute them  through the Internet. Data is only transmitted by satellites during their passing of ground stations, with an average of one track per day, every 6-8 minutes. We have deployed many ground stations, mainly in Changsha and Xinjiang. The ground stations also serve as gateways to the cloud computing platform. All of these together form Tiansuan Constellation.

\subsection{Operation Mechanism}
The constellation operates according to the open principle. For research institutes, it provides a platform for scientific research. Institutions can join Tiansuan constellation  either by submitting research requirements or by enhancing the constellation in collaboration. For satellite companies, Tiansuan constellation is a complement to their ecosystem. Most companies find it inconvenient to complete some research work for profit reasons. Tiansuan constellation fills this gap by gathering researchers to work on cutting-edge exploratory problems. As a result, the overall cost for companies and research institutes can be reduced to better stimulate industry development. Other organizations can take advantage of Tiansuan constellation for testing   new devices or payloads. 
\section{Potential Spectrum of Experiments }

 It is difficult to envision the full spectrum of experiments
creative researchers may run using Tiansuan constellation, several main
 research directions that it would support are as follows.
 
 \subsection{ Networking }
The ambitious plan for providing the Internet through satellite constellations has attracted the attention of both industry and academia \cite{LarLEO}.
Satellite communication, as an enabling technology of 6G networks, is crucial to achieving global coverage of mobile networks.
Tiansuan constellation can support networking-related experiments at the physical layer (e.g., link measurement),  network layer (e.g., routing), transport layer (e.g., congestion control), and application layer (i.e., interactive applications). Besides, it can also support experiments on the network control plane (e.g., core networks).

For further clarification of the preceding assertion, we elaborate on one of them, the deployment of 6G core networks. With the rapid construction of space Internet infrastructure, traditional ways of managing constellations cannot keep up. Satellites are instead made from generic hardware and deployed using custom software. By doing so, the onboard network, computation, and storage resources can be utilized more efficiently. Core networks connect heterogeneous access networks and data networks. The combination of satellite constellations and next-generation core networks will lead to a variety of application scenarios. As an example, LEO satellites are capable of delivering Internet services with lower latency over long distances \cite{bhattacherjee2018gearing} \cite{handley2018delay}. Core networks deployed on satellites will reduce the latency of signaling interactions incurred by the control plane, enabling much more real-time services. On the other hand, satellite constellations can enhance their capabilities with the next-generation core networks. Satellite constellations are becoming increasingly difficult to manage due to their explosive growth. By deploying lightweight core networks on satellites, constellation management can be more flexible and agile.

\begin{figure}[t]
\begin {center}
\centerline{\includegraphics[width=0.4\textwidth]{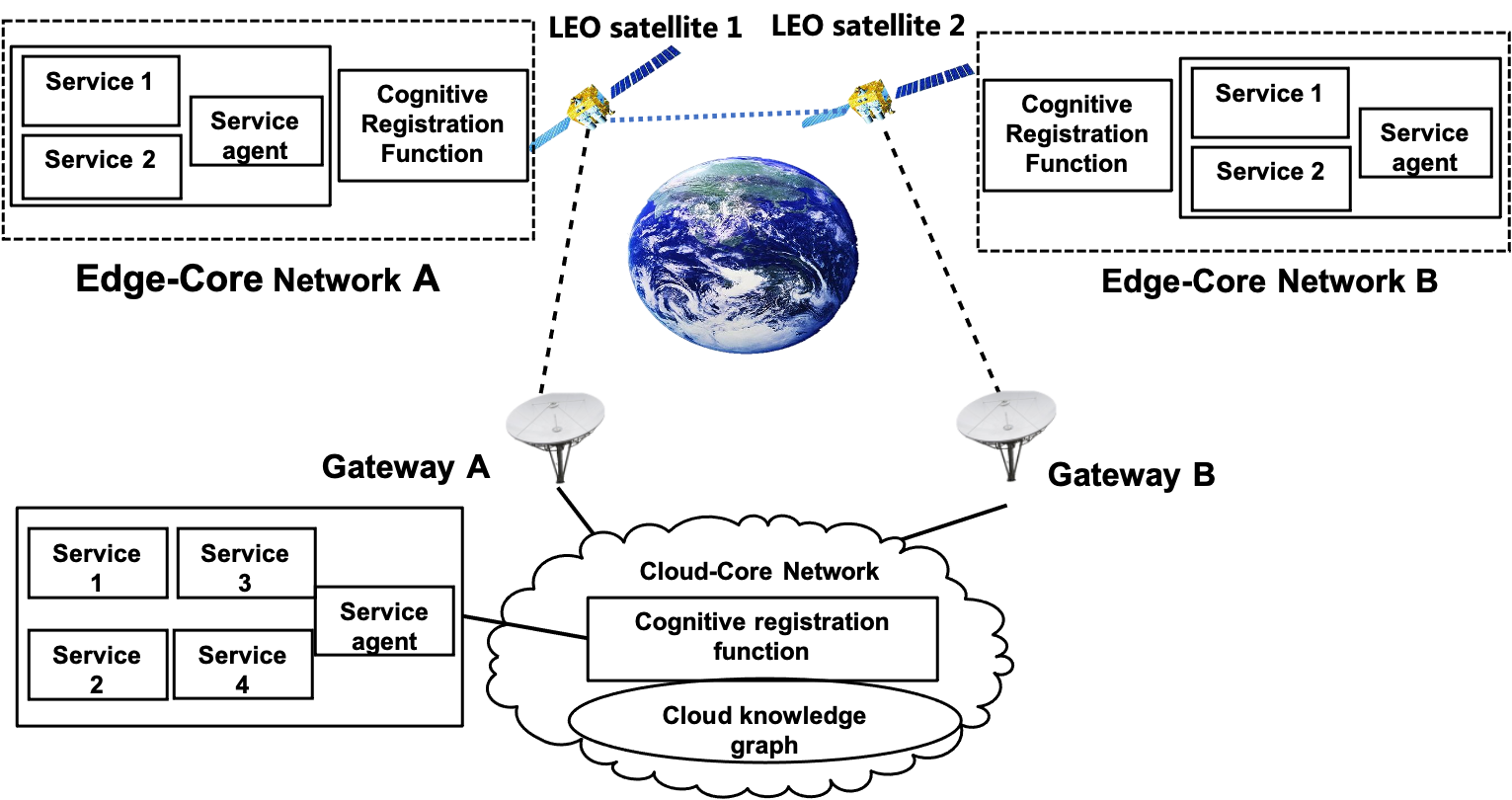}}
\caption{ Cognitive Service Architecture of 6G core network.}
\label{fig: 6G}
\end {center}
\vspace{-6ex}
\end{figure}

Developing the core network functions for LEO satellite nodes is the key to integrating the 6G core networks with satellite constellations. However, there are two challenges to overcome. First of all, satellite networks are experiencing rapid changes in network topology and connectivity. Terrestrial mobile core networks should virtualize their functions more adaptively in space. Moreover, power supply and processing resources on satellites are limited. Network functions and applications should be orchestrated more efficiently. Thus, one of the key objectives of the Tiansuan constellation is to verify the cognitive service architecture for 6G core networks as shown in Fig.~\ref{fig: 6G}.

\subsubsection{Mobility Management}
Satellites in highly elliptical orbits will have a long latency. While LEO satellites can reduce latency significantly, they still suffer from short link duration.

Even though a large ground-orbiting satellite constellation can solve the short link duration problem, frequent handovers will affect the quality of the service. Thus, the mobility management of the integrated satellite-terrestrial network is an important experiment of the Tiansuan constellation.

\subsubsection{Core Network Coordination}
Inspired by the distributed nervous system of the octopus, we propose a cognitive service architecture for the 6G core network. There is one difference between cognitive service architecture and the previous core network architecture: the core network is divided into the edge core network and the cloud network. The edge core networks are similar to the peripheral nervous system of an octopus's arms, as they provide the majority of the network control. The cloud core networks, acting as the central brain of octopus, are only responsible for guiding and assisting the coordination between edge core networks. In addition to satellite communication, the edge core network will be deployed on satellites. With the help of the Tiansuan constellation, we will deploy edge core networks and evaluate their performance.

\subsubsection{Serverless and Stateless Design}

Serverless and stateless design is an effective way to achieve lightweight deployment. With serverless, the original network element is split into several independent functions. Each functionality can be independently developed, tested, deployed, and evolved. Different functionalities can call each other through flexible interfaces. Stateless separates status from functions for network elements. This will benefit satellite network reconfiguration and synchronization.

In an integrated space-air-ground network, the limited resources and weak capabilities of space-based nodes must be considered. The 6G core network can be implemented on the satellite with more efficient resource usage by utilizing serverless network functions. By using the remote interface, network functions can access abundant resources on the ground. This allows us to deploy lightweight applications without a heavy network component. By using this design, network functions can be deployed in different locations according to business needs and resource availability. In addition, it also supports network function cancellation and redeployment.

\subsubsection{Knowledge Graph}

Satellite networks are highly dynamic and have uncertain links. It is therefore more difficult to select a reasonable node and plan an optimal flow path for users. Creating a knowledge map of users, networks, and services can provide an overview of their real-time status and logical relationship. In addition, it provides real-time user plane node selection and dynamic traffic path planning for satellite networks.

The control plane of onboard core networks controls access and mobility, as well as establishes and schedules user sessions. With the uncertainty of the network, large-scale user access, and the dynamics of satellite-to-ground services, the control plane signaling process has become increasingly complex. Through its knowledge extraction and reasoning capabilities, the knowledge graph can support decision-making processes in the network. Therefore, it simplifies signaling interactions in the network.

Nodes and attributes are the basic elements of the knowledge graph. When the collected information has temporal characteristics, it can be used to describe the state of the nodes. The mobility of satellite networks and the need to ensure service continuity necessitate frequent network element migration. The characteristics of the knowledge graph can be used to plan reasonable migration strategies for network element functions. As a result, the knowledge graph ensures better service continuity.

  \subsection{Computing}
 LEO satellite constellations  promise to provide  global, low-latency, high-bandwidth Internet service, which has generated tremendous interest in both  industry and the research community. There is also a potential opportunity that constellations offer computing services wherever people  want.   Motivated by  edge computing on the ground, satellite edge computing   has been proposed  by placing computing resources at the LEO satellite constellation \cite{10.1145/3373376.3378473,9444334}.
 Tiansuan constellation would  support  satellite-borne computing platform,  constellation resource management, general computing service, and heterogeneous satellite data integration, which could provide technical reference for supporting the construction of space data centers.
  In the following, we provide more details on two topics of processing space-native data, namely real-time earth observation and AI in space. Besides, we also discuss how to provide general computing services for ground users.

 \subsubsection{Real-Time Earth Observation}
About 45\% LEO satellites in orbit  are used for earth observation, which can be further used in various applications, such as disease spreads \cite{Bhattachan2019UsingGD}, crop monitoring\cite{Aragon2018CubeSatsEH}, and nature disaster management \cite{Barmpoutis2020ARO} (e.g. forest fires, floods, etc.).  As the constellation size increases,  the conflict between large space-native data volume  and limited downlink capacity  is the bottleneck of real-time Earth observation. There are two potential ways to address the challenge,  (1) modeling  the freshness of the interested information and only downloading   freshest information,  (2) leveraging multiple  distributed  ground stations to schedule downloads effectively.

   \textbf{Freshness of Information.}
      Since the real-time applications need fresh information for the decision-making, it is critical to make sure that the received information is valuable and timely. Thus, one of the main reasons behind the difficulty in the deployment of various real-time earth observation applications is the hardness of obtaining fresh data. Unlike traditional terrestrial transmission, data must wait on the satellite before it establishes a link to the ground station. This adds the staleness of data received by the ground stations. Furthermore, when the constellation size is small, the ground stations have very limited time to establish links with satellites, it is very difficult for them to download so much high-quality image data in such a limited time, not to mention the bottle-necked bandwidth and contention. Age of information (AoI) \cite{Kaul2012RealtimeSH,Kaul2011MinimizingAO} as a metric of capturing the freshness from the destination widely used in time-sensitive applications. This motivates us to conduct AoI-aware research on the data sampling, transmission, and processing on the Tiansuan constellation project for improving the performance of earth observation systems.  
   
   \textbf{Ground Station Deployment.}
 The   current centralized ground stations incur high deployment  costs \cite{vasisht2021l2d2},  large  delay jitter\cite{kassing2020exploring}, and weak scalability. In general, the cost of deploying a ground station is around a million dollars, primarily due to the current demand for professional equipment and maintenance \cite{vasisht2021l2d2}. For new entrants, the cost of permitting and establishing ground stations is prohibitive.
To meet the specific delay requirements of different users at different times, we could study a suitable ground station deployment strategy. The development of lightweight and scalable ground stations will also be non-trivial to realize the interconnection and intercommunication of the satellite-to-ground network.
With the large number of satellites and ground stations, we must dynamically schedule ground station-satellite links while taking into account orbits, the quality of the link, and the weather, etc. By  dynamic ground station-satellite link planning, congestion is avoided and transmission efficiency is improved. Forecasting is considered a good way to solve this problem \cite{vasisht2021l2d2}. Traditional traffic forecasting methods, however, have been unable to accurately predict traffic on satellite networks due to the spatial and temporal correlation characteristics. Spatial-temporal predictions based on the impacts of past workloads themselves (temporal) and their mutual impacts (spatial) have received considerable attention in recent years, mainly focused on traffic flow predictions\cite{lin2021rest, wang2020traffic,li2018dcrnn_traffic}. The satellite downlink rate varies with the conditions of the channel at a given location. By using historical data and environmental information (weather conditions) to create a satellite link spatial-temporal prediction model, we can analyze the mutual influence between different links and the autocorrelation over time.
      \subsubsection{AI in Space}
      The rapid development of satellite edge computing and satellite Internet will promote the popularity of on-board applications, which eventually enable  satellite-based AI techniques.     
      
      \textbf{Federated Learning in Space.}
       Federated learning (FL) \cite{mcmahan2017communication,li2018federated,bonawitz2017practical}, as a distributed learning paradigm, has great prospects for widespread deployment in satellites.   Traditional image detection methods need to transmit these images to the ground station for recognition. But these methods face two key challenges:   (1) The uplink and downlink bandwidth is limited, and the transmission process has a large latency, and  (2) the transmission process is relatively fragile, and the transmission may be interrupted, the data may be lost. In addition,  privacy protection is also a key requirement for satellite applications. By FL, we can directly process and analyze the collected information on the satellite, while protecting the satellite's data privacy \cite{yurochkin2019bayesian}. We propose to establish a federated learning and AI acceleration platform based on Tiansuan constellation, through which we further study and verify the following research points:  (1) machine learning model training and acceleration capability verification based on space-borne computing equipment, (2) design and verification of AI algorithms for specific satellite application scenarios, and  (3) establishment and verification of federated learning experimental platform based on Tiansuan constellation.
   
  Meanwhile,  FL, just like traditional deep learning systems \cite{pei2017deepxplore,tian2018deeptest}, often demonstrates incorrect or unexpected corner-case behaviors, especially in the harsher space environment. We can design a systematic testing tool for automatically detecting erroneous behaviors of FL-driven satellites that can potentially lead to data invalidation analysis \cite{yang2021advances}. Our tool is designed to automatically generate test cases leveraging real-world changes in the space environment like meteorites, lighting conditions, shooting angle, etc. We can further show that the test inputs generated by our testing tool can also be used to retrain the corresponding FL model to improve the model’s robustness.

  \textbf{Inference in Space.}
As mentioned above, offloading all deep learning tasks to the ground station  is one of the current mainstream research methods \cite{del2021board, tang2021computation}.
However, the challenge is the limited bandwidth in practical applications. It's reported that about two-thirds of the earth’s surface is covered by clouds at any time
and satellites often capture and save a large number of useless images,  which cause the downlink bandwidth occupied and great uncertainty in network environment.
The current experimental satellites have airborne AI processing capabilities.
It's possible to leverage on-board inference to identify and remove useless images and only  send useful ones to the ground station for subsequent processing. 
Based on a variety of satellite SoCs, using the characteristics of heterogeneous computing platforms to accelerate AI tasks is worth expecting \cite{rapuano2021fpga}.
Moreover, speculative inference can benefit data-driven applications that involve the fusion of multi-modal satellite signals \cite{ochoamachine}. 
Real-time actuation can benefit from speculative inference. For instance, satellites navigation or grasping can leverage multi-modal signals from different points 
to finish difficult tasks in complex environments.

\subsubsection{Space Service Computing}
   The ground-satellite connectivity and inter-satellite connectivity are time-varying due to the dynamics of the satellite network topology. 
  In such a dynamic context, (1) how to select the ground-satellite path and inter-satellite path and provide reliable network service, (2) how to realize the dynamic service deployment, computing offloading, service migration and service coordination are still to be explored.
Micro-service and function as a service are the lightweight application architectures for services in satellites. They decompose applications to finer-grained service components which can be deployed and executed quickly and independently. However, (1) how to partition applications to a set of micro-services or functions, (2) how to deploy dependent micro-services and functions dynamically, (3) how to register and manage the highly distributed micro-services and functions, and (4) how to allocate resource for these large-scale micro-services and functions are still to be researched.

  \subsubsection{Cluster Orchestration}
    Satellite edge computing can realize computation offloading and process the computation tasks at LEO satellites, which can save satellite-ground or inter-satellite link transmission bandwidth and reduce the impact of large satellite-ground link delays. There are some issues to be addressed. First, the computing power of a single LEO satellite is limited, then how to select satellite clusters to process the task cooperatively.  Second, how to design a cooperation strategy to adapt to 
LEO satellites' high mobility, and wide-area load imbalance. Third,  how to distribute the computation tasks to realize the balance between  multiple objectives, such as   on-board energy efficiency, on-board computing delay, communication transmission delay. 
The demand of computing resources and satellite-ground collaborative AI inference can be used to optimize multi-satellite node computing task allocation, computing data migration, inter-satellite signaling, and computing result return strategies.

Besides, it is necessary to manage a large number of satellites in the  constellation.
Traditional cluster orchestration technology has been widely adopted to manage a huge number of computation and storage resources in the network~\cite{DBLP:conf/nsdi/HindmanKZGJKSS10,DBLP:conf/eurosys/VermaPKOTW15,DBLP:conf/usenix/KaranasosRCDCFH15,DBLP:conf/usenix/DelgadoDKZ15}.
In recent years, with the continuous development of edge computing~\cite{DBLP:conf/imc/XuFMZLQWLYL21}, 
integrating the emerging edge nodes and platforms to cluster management provides a huge number of benefits, such as resources utilization improvement, the software deployment process simplification and the compatibility with cloud-native software stack~\cite{DBLP:conf/osdi/TangYVKMKACGCCG20,DBLP:journals/cacm/BurnsGOBW16}.
Treating LEO satellites as edge nodes gives a unified way to schedule and deploy the applications on the satellite by reusing the existing applications. This approach avoids re-developing the platform-specific programs, while moving toward the vision of \textit{software-defined satellite computing}.
Previous work has investigated the difficulty of deploying traditional container orchestration technology on the satellite co-located with the ground station~\cite{DBLP:conf/hotedge/BhosaleBG20}, where we call this inter-satellite orchestration.
However, there may be a bunch of computing units in a single satellite (e.g., Raspberry Pi). 
Scheduling tasks to these computing units efficiently can fully utilize the computing resources on the satellite.
Inspired by the inter-satellite orchestration, we can also use this approach to integrate computing units inside the satellite and provide services (e.g., KubeEdge~\cite{DBLP:conf/edge/XiongSXH18} provides huge advantages in managing edge nodes).
We call this inner-satellite orchestration.
By using this multi-level management technology, we think the computing efficiency can achieve a great enhancement.

 \subsection{Satellite Operation System }
 In order to adapt to the increasing demand for satellite software~\cite{DBLP:journals/corr/Khan15b,DBLP:conf/hotedge/BhosaleBG20}, it is urgent to design and develop an underlying operating system for the satellite Internet.  
This new operating system features miniaturization, ubiquity and universalization, and will be installed and verified on the orbiting satellite, so as to build an open satellite software environment.
Due to the lack of software and API in the traditional operating system ecosystem~\cite{stankovic2004real}, the satellite operating system should be a soft real-time and hard real-time operating system for embedded hardware. 
Therefore, the satellite operating system needs to support users' definable and quantifiable multi-task real-time requirements, such as databases, machine learning engines, image and video processing.
However, the implementation may face a list of challenges: 
(1) How to propose an approach to upload third-party software while guarding the security and extensibility? This requires the operating system to support a variety of technologies, such as static code analysis, run-time stain analysis, and execution sandbox. 
(2) How to ensure memory safety during software runtime? We can leverage the Rust programming language to implement the operating system for maximum memory safety~\cite{levy2015ownership,lankes2019exploring}. We also need to improve the relevant toolchain and system middleware to pave the way for future development. 
(3) How to ensure the reliability of the operating system in the extreme environment of space (e.g., single event ﬂipping)? 
This requires the developers of the operating system to research on multi-machine backup and fault-tolerance technology to ensure the stability of the software and the system.

  \subsection{Security and Reliability}
The security  and reliability of satellite constellations are fundamental to  enable its  great potential.
 With the advantage of wide coverage of satellites, everyone in the world is allowed to participate in the network while there may be some security issues.
 Besides, in the LEO environment, due to the combined effects of temperature and radiation, it is very challenging to ensure the reliability of LEO micro-satellite service computation. Tiansuan constellation would support solutions that guarantee security and reliability.
   \subsubsection{Security of Constellation}
Blockchain is a decentralized data storage technology with ensuring the data transparency, integrity, and immutability.
The feasibility of blockchain deployment on satellites is indicated by previous work. SpaceChain $\footnote{https://spacechain.com/}$ employs blockchain on satellites to provide secure orbit data storage, reducing the dependency of blockchain on the ground network. Two nanosatellite-based blockchain nodes have been launched into orbit aboard Chinese Long March rockets in 2018. A successful test of space-to-ground blockchain transactions has been done by SpaceChain in 2019. Blockstream$\footnote{https://blockstream.com/}$  has launched the service of Blockstream satellite aimed at broadcasting the Bitcoin blockchain to the entire planet via satellite. Mital et al. demonstrated the usage of blockchain’s smart contract and distributed ledger for multi-sensor satellite by employing software STK \cite{ mital2018blockchain }. 

While the work mentioned above demonstrates their implementation of blockchain on satellites, there are remaining issues to be studied. As satellites are the main source of spatial data, the storage of data on satellites and the transmission to the ground is one of the most crucial issues. Digital signature on the satellite is required to be researched to provide the traceability of data so as to ensure immutable storage. The satellite-ground collaborative blockchain transmission is a promising solution to guarantee the reliability of links and the credibility of data. The limited resource and the special architecture of satellite networks increase the difficulty of problem modeling and analysis. Moreover, simulations on the ground are insufficient to conduct further researches. Fortunately, the Tiansuan satellite as an open-source platform provides the opportunity for real experiments, which would speed up the evolution of blockchain deployment on satellites.
      
 \subsubsection{Reliability of Constellation}

Benefiting from the wide application of commercial-off-the-shelf  in LEO satellites, the cost of LEO satellites has been further reduced, and at the same time, LEO satellites are also endowed with excellent application compatibility.
However, the commercial-off-the-shelf  is not designed for the harsh environment of space.
In the LEO environment, since there is almost no atmosphere, satellites mainly rely on thermal radiation for heat dissipation. In the illuminated area, the temperature can rise to ~400K, while in the non-illuminated area it will drop to ~150K \cite{guide}. The low-mass LEO satellites frequently enter and exit the illuminated area, causing their working temperature to fluctuate  drastically and frequently. Besides, affected by the energy distribution strategy and overheating protection strategy on the satellite, the computing power of the satellite will fluctuate greatly too.
On the other hand, high-energy charged particles in space are affected by the earth's magnetic field, forming  the Van Allen Belts. The commercial-off-the-shelf  components within its range may have a single-event effect \cite{ISO21980} at any time, which can lead to calculation errors, temporary, or even permanent failures of the components.
Traditional reliability assurance strategies, such as the introduction of redundancy, checkpoints, etc., are difficult to apply to nowadays more complex and large-scale satellite service instances on satellites with extremely limited energy reserves and computing power.
Therefore, the research on the service calculation guarantee strategy of LEO satellites is a key point in the Tiansuan satellite constellation project.

  \subsection{Hardware Testing}
The Tiansuan constellation also supports space experiments and in-orbit testing of new hardware devices. For example, the performance of sensors, cameras, and antennas can be tested on Tiansuan constellation. Besides, computing acceleration equipment such as acceleration boards can also be deployed to verify its performance and 
reliability in the space environment. Furthermore, the Tiansuan constellation can also carry out the distributed deployment of the computing platform to evaluate its performance in the satellite environment.


\section{Case Study}
We  have deployed two use cases in  both two experimental satellites  of Tiansuan constellation named Baonyun and Chuangxingleishen, which will be launched on December 2, 2021.
\subsection{Satellite-borne B5G Core Network}
  \begin{figure}[t]
\begin {center}
\centerline{\includegraphics[width=0.45\textwidth]{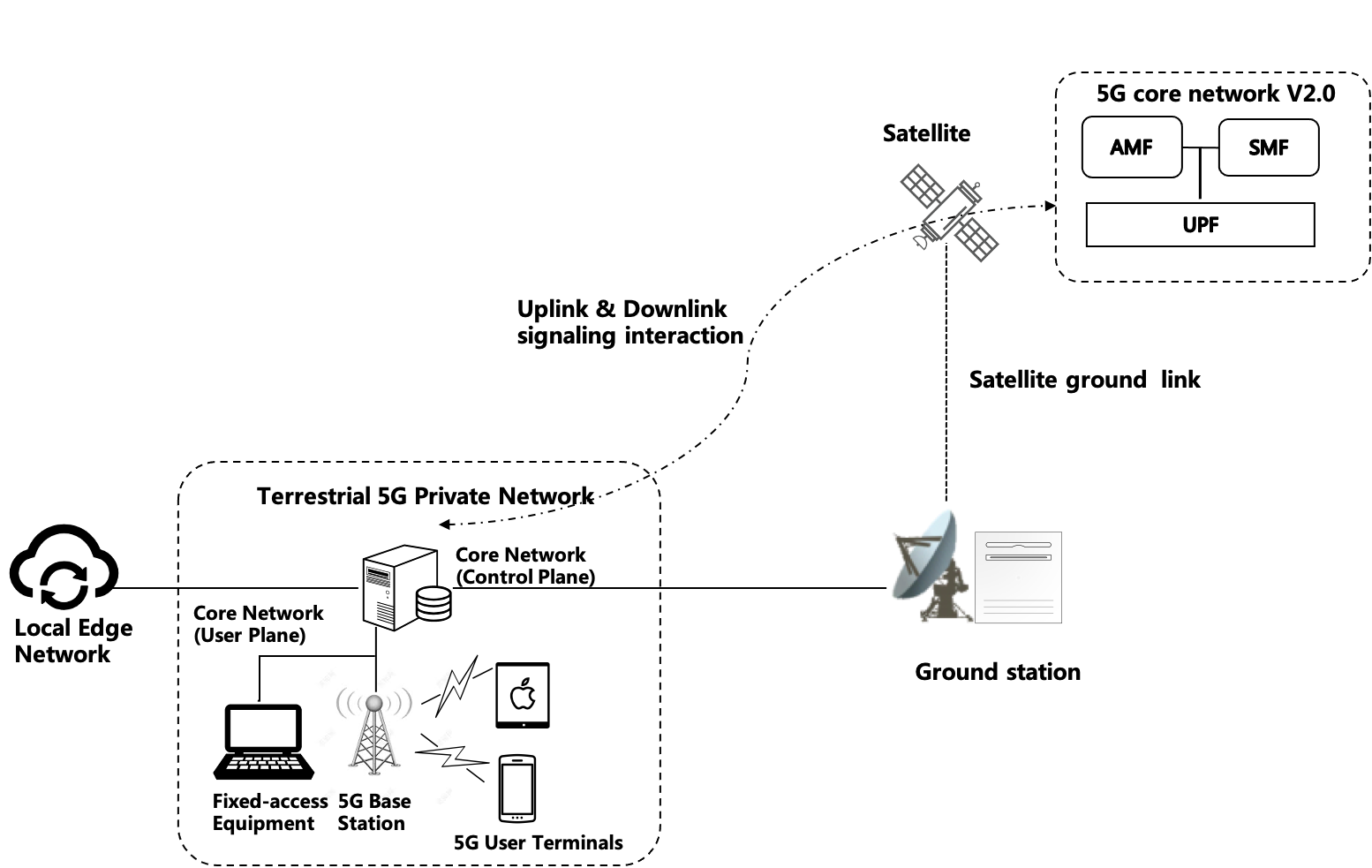}}
\caption{ Satellite-Borne B5G Core Network.}
\label{Fig:5Gcore}
\end {center}
\vspace{-6ex}
\end{figure}
On August 9, 2021, we conducted the first  deployment test of a 5G core network system on the satellite called TY20. We tested the signaling interaction (base station and UPF) between the on-board B5G core network and the terrestrial private 5G network in both the control plane and the user plane. Satellite control commands were sent through the uplink telemetry link. The B5G core network on the satellite successfully realized user registration, session establishment, and control of satellite equipment. Downlink telemetry monitoring showed that the three main functional components of the core network were operating normally. It also showed that the core network control data was generated correctly. The control data generated by the B5G core network was downloaded to the terrestrial private 5G network. In this way, we realized the local offloading of edge computing controlled by satellites, and conducted tests such as video calls. As shown in Fig. 3, we have successfully deployed a lightweight B5G core network on both the Baonyun and Chuangxingleishen satellites. This core network is the updated version of the former 5G core network. It enhances the signaling interactions and can be used to set up video calls based on Session Initiation Protocol. Tests of functionalities and performance will be conducted after the two satellites are launched.

 \subsection{Image Inference Based on KubeEdge}
\begin{figure}[t]
\begin {center}
\centerline{\includegraphics[width=0.45\textwidth]{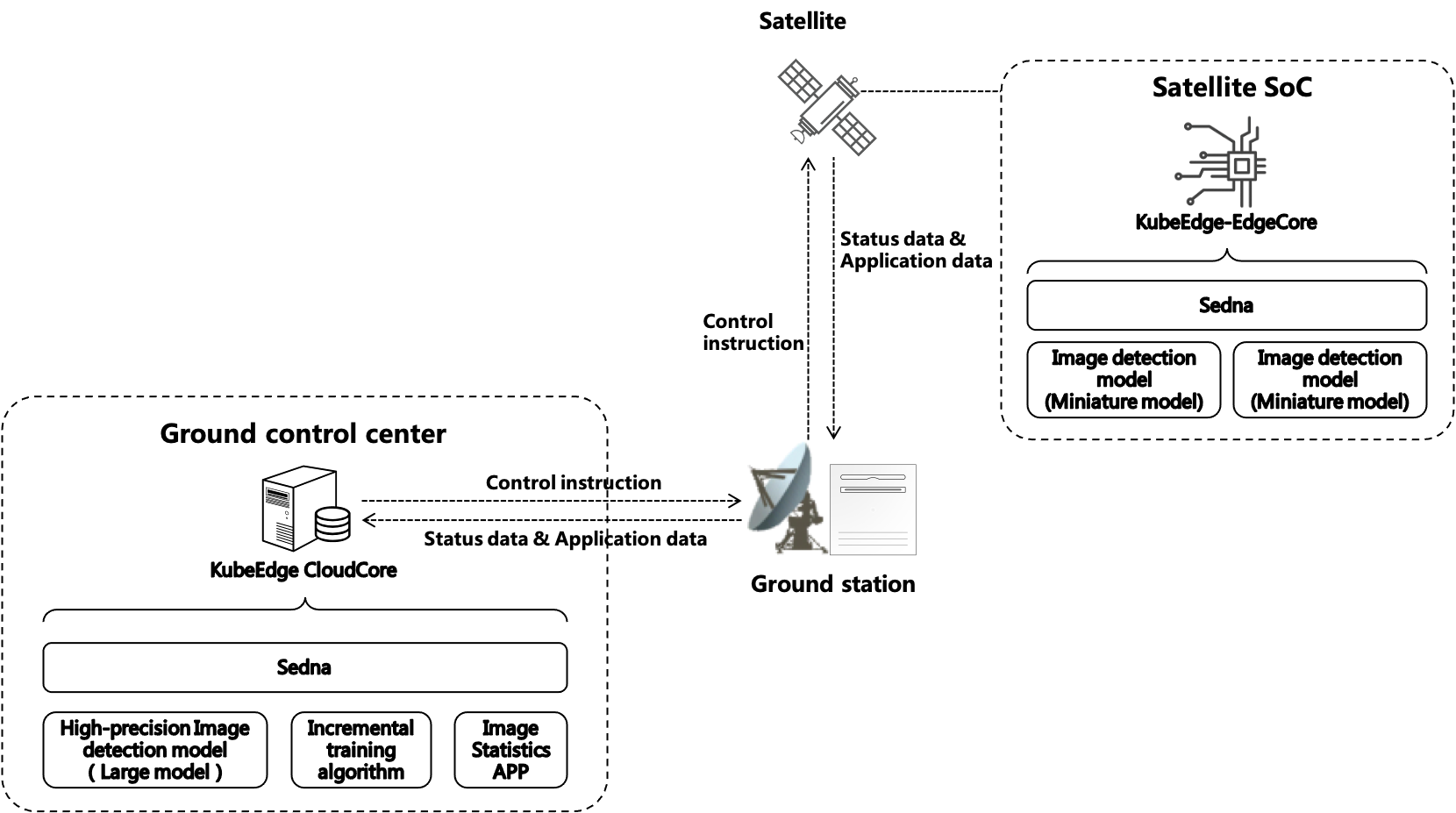}}
\caption{ Image Inference based on KubeEdge.}
\label{Fig:cubedge}
\end {center}
\vspace{-6ex}
\end{figure}

We have implemented KubeEdge and the corresponding AI-related system on both Baonyun and Chuangxingleishen satellites as shown in Fig.~\ref{Fig:cubedge}. 
In this system, we deployed a central controller using a Linux server on the ground. 
The satellite establishes an intermittent connection with the central controller inside the KubeEdge runtime according to the position of the satellite. 
Based on Sedna, the KubeEdge AI extension, we deployed two image detection models, a lightweight model and a high-precision model on the satellite and the central controller respectively. 
During the inference of the AI pipeline, the satellite will capture the image and detect whether it is an object of interest.  
When the conﬁdence of inference is high, the satellite will use this concrete result for later processing. 
Only when the conﬁdence is low, the satellite will transfer the image to the central controller on the ground, aiming to get an exact inference result using the high-precision image detection model. 
Such a collaborative approach utilizes the computing resources on the satellite, which reduces the end-to-end latency in an AI pipeline. 
\section{Conclusion}
In this paper, we introduce  a global research platform, Tiansuan constellation, to enable experiment evaluations on real satellite constellations. We present the design of Tiansuan constellation  in detail and state how can different parties  use  our Tiansuan constellation. Moreover, we list several research topics that Tiansuan would enable and provide two case studies of real deployment. We hope that  more institutes and individuals
interested in LEO satellites can join us in this exciting research area. 
\bibliographystyle{IEEEtran}
\bibliography{IEEEabrv,reference211125}

\end{document}